\documentstyle[11pt]{article}
\def\fnote#1#2{\begingroup\def\thefootnote{#1}\footnote{#2}\addtocounter
{footnote}{-1}\endgroup}

\begin{document}

\hfill{UTTG-09-20}

\vspace{20pt}

\begin{center}
{\large {\bf { Massless Particles in Higher Dimensions}}}

\vspace{20pt}

Steven Weinberg\fnote{*}{Electronic address:
weinberg@physics.utexas.edu}\\
{\em Theory Group, Department of Physics, University of
Texas\\
Austin, TX, 78712}

\vspace{30pt}

\noindent
{\bf Abstract}
\end{center}

In spacetimes of any dimensionality, the massless particle states that can be created and destroyed by a field in a given representation of the Lorentz group are severely constrained by the condition that the  invariant Abelian subgroup of the little group must leave these states invariant.  A number of examples are given of the massless one-particle states that can be described by various tensor and spinor-tensor  fields, and a speculation is offered for the general case.

\vfill

\pagebreak

\begin{center}
{\bf 1. INTRODUCTION}
\end{center}

For a century  physicists have speculated that our familiar four-dimensional spacetime  may really be embedded in a higher dimensional continuum [1].    It is simplest and most usual to suppose that this continuum is a $d$ dimensional spacetime in which in locally inertial frames the laws of nature are invariant under the Lorentz group $SO(d-1,1)$.  
If a general $SO(d-1,1)$ Lorentz transformation $\Lambda^\mu{}_\nu$ acts on physical states as a unitary  representation $U(\Lambda)$  of $SO(d-1,1)$, then fields are characterized by their transformation
\begin{equation}
U(\Lambda) \psi^n(x) U^{-1}(\Lambda)=\sum_m D^{nm}(\Lambda^{-1})\psi^m(\Lambda x)
\end{equation}
where $D^{nm}(\Lambda)$ is a finite-dimensional generally non-unitary matrix representation of $SO(d-1,1)$.
As Wigner [2] first pointed out for the case $d=4$, one-particle states must be characterized as representations of the little group, the subgroup of $SO(d-1,1)$ that leaves some standard spatial momentum ${\bf k}$ invariant.  That is, for any Lorentz transformation $W^\mu{}_\nu$ for which $W^\mu{}_\nu k^\nu =k^\mu$ the states $|{\bf k},\sigma\rangle$ with standard momentum satisfy
\begin{equation}
U(W)|{\bf k},\sigma\rangle=\sum_{\bar{\sigma}}d_{\bar{\sigma},\sigma}(W)|{\bf k},\bar{\sigma}\rangle
\end{equation}
where $d_{\bar{\sigma},\sigma}(W)$ is a unitary matrix representation of the little group.
The question then arises, what sort of representations of the little group can arise for particles described by a field  that transforms according to some given  representation of $SO(d-1,1)$?

For massive particles the answer is easy.  The standard $k^\mu$  can be taken to have spatial components ${\bf k}=0$, and the little group is then the semi-simple rotation group   $SO(d-1)$.  Particles described by a given field can furnish any representation of this  $SO(d-1)$ subgroup that is contained in the representation of $SO(d-1,1)$ furnished by the field.

The case of massless particles presents some complications.  This case is of special interest, since from the perspective of the very high energies where higher dimensional theories might be relevant, all of the particles we observe are massless.  Finding what sort of  massless particle can be described by various types of field in higher dimensions is complicated by the fact that  their states must be classified as representations of the little group that leaves invariant some  standard {\em non-zero}  $d-1$-momentum and, unlike the massive case, this little group contains an invariant Abelian subgroup.  To avoid the introduction of new continuous conserved quantities, particle states must be invariant under this subgroup.  For any spacetime dimensionality $d$  the massless particle states must  furnish a representation of the remaining semi-simple subgroup $SO(d-2)$ of the little group.  But not all representations of $SO(d-2)$ that are contained within the representation of $SO(d-1,1)$ that is furnished by the field are consistent with the requirement that the invariant Abelian subgroup is represented trivially.  So it is not immediately obvious  how  massless particles described by a given field can transform under $SO(d-2)$.  This is the problem addressed in this paper.

To put  this problem in greater detail, we consider the matrix element
\begin{equation}
u_\sigma^n\equiv \langle {\rm vacuum}|\psi^n(0)|{\bf k},\sigma\rangle\;.
\end{equation}
Inserting Eqs.~(1) and (2) gives the requirement
\begin{equation}
\sum_{\bar{\sigma}} d_{\sigma,\bar{\sigma}}(W)\,u_{\bar{\sigma}}^n=\sum_{m}D^{nm}(W)u^m_{\sigma}\;,
\end{equation}
for any element $W$ of the little group.    (We have used $d^*_{\bar{\sigma},\sigma}(W^{-1})=d_{\sigma,\bar{\sigma}}(W)$.)  Our question is, for a given representation $D(\Lambda)$ of the Lorentz group, what representations of $SO(d-2)$ allow  a non-zero matrix element $u_\sigma^n$ for which $d_{\bar{\sigma},\sigma}(W)=\delta_{\bar{\sigma},\sigma}$ for any $W$ in the invariant Abelian subalgebra of the little group?  
As a by-product of answering this question, we will find field equations that are satisfied by  the matrix elements
\begin{equation}
\langle {\rm vacuum}|\psi^n(x)|{\bf k},\sigma\rangle=e^{ik\cdot x}u_\sigma^n\;.
\end{equation}

So far we have been implicitly  taking the states $|{\bf k},\sigma\rangle$ to be eigenstates of the full Hamiltonian, and the field $\psi^n(x)$ to be an interacting field.  But the mathematical requirement on $u^n_\sigma$ is the same if we take $|{\bf k},\sigma\rangle$ to be eigenstates of the free-particle Hamiltonian, and $\psi^n(x)$ to be a free field.  In this case, the  $u^n_\sigma$ can be used to construct the free field:
\begin{equation}
\psi^n(x)=\int d^{d-1}p\sum_\sigma\left[u^n({\bf p},\sigma)a({\bf p},\sigma)e^{ip\cdot x}+
v^n({\bf p},\sigma)b^\dagger({\bf p},\sigma)e^{-ip\cdot x}\right]\;,
\end{equation}
where $a({\bf p},\sigma)$ are the annihilation operators for the massless one-particle state; $b({\bf p},\sigma)$ are the annihilation operators for the corresponding antiparticle;
the coefficient function $u^n$ is given by
\begin{equation}
u^n({\bf p},\sigma)=\sqrt{\frac{k^0}{p^0}}\sum_m D^{nm}\Big(L({\bf p})\Big)u_\sigma^m\;,
\end{equation}
where $L({\bf p})$ is the $SO(d-1,1)$ transformation that takes the standard momentum ${\bf k}$ to ${\bf p}$; and likewise for antiparticles [3].  In this context our question is, what kind of massless  particles characterized by their transformation under $SO(d-2)$ can be used to construct a free field characterized by its transformation under $SO(d-1,1)$?
The free fields constructed in this way satisfy the same field equations as will turn out to be  satisfied by the matrix element (5) of interacting fields.

The answer to our question is well known for $d=4$ [4].  The finite-dimensional (and in general non-unitary) irreducible representations of $SO(3,1)$ are characterized by a pair $A$ and $B$ of integers and/or half-integers.
These are defined by ${\bf A
}^2=A(A+1)$ and ${\bf B
}^2=B(B+1)$, where $$A_a\equiv \frac{1}{4}\sum_{bc}\epsilon_{abc}J_{bc}+\frac{i}{2}J_{a0}\;,~~~~B_a\equiv \frac{1}{4}\sum_{bc}\epsilon_{abc}J_{bc}-\frac{i}{2}J_{a0}\;,$$ where here $a,\,b,\,c$ run over the values $1,\,2,\,3$ and $J_{ab}=-J_{ba}$ and $J_{a0}=-J_{0a}$ are the generators of $SO(3,1)$.  For $d=4$  the representations of the semi-simple part $SO(2)$ of the little group are one-dimensional and characterized by a number, the helicity.  The massless particle described by a field of type $(A,B)$ can only have helicity 
\begin{equation}
\lambda=B-A
\end{equation}
For $A\geq B$ (or $B\geq A$) the free field and the matrix element (5) of the interacting field are spacetime derivatives of order $2B$ (or $2A$) of a free field (or field matrix element) of type $(A-B,0)$ (or $(0,B-A)$).

Here are  results for spacetimes of general dimensionality $d$, derived in the following sections:
\begin{itemize}
  \item A symmetric traceless tensor field of any rank can  have non-vanishing matrix elements (3) only for a massless particle that transforms trivially under $SO(d-2)$.  Further,  the matrix element (5) of a symmetric traceless tensor field of rank $N$ is necessarily the $N$-th spacetime partial derivative of a scalar, as is the free field.  This is the generalization of the result for $d=4$, that a symmetric traceless tensor field of rank $N$ has Lorentz transformation of type $(A,B)$ with $A=B=N/2$, and therefore can only describe particles of helicity $B-A=0$.
  \item A completely antisymmetric tensor field of rank $N< d $ can only have non-vanishing matrix elements (3) for a massless particle that transforms  under $SO(d-2)$ as an antisymmetric tensor of rank $N-1$.  The matrix element (5)  (and the free field) is necessarily the exterior derivative of an antisymmetric potential of rank $N-1$, and therefore has vanishing exterior derivative.  This is the generalization of the      familiar result for $d=4$, that an antisymmetric  tensor field of rank 2 has Lorentz transformation of type $(1,0)\oplus (0,1)$, and therefore can only describe particles of helicity $\pm 1$, and is the curl of a vector potential.
  \item A rank 4 traceless tensor field with the symmetry property of the Weyl tensor (the traceless part of the Riemann-Christoffel curvature tensor) can have non-vanishing matrix elements (3) for a massless particle that transforms under $SO(d-2)$ only as  a second-rank symmetric traceless tensor.  The matrix element (5)  (and the free field) is necessarily the antisymmetrized second derivative of a symmetric second rank potential.  This is the generalization of the result for $d=4$ that the Weyl field has Lorentz transformation of type $(2,0)\oplus (0,2)$ and therefore can only describe massless particles with helicity $\pm 2$.   
  \item A field with the Lorentz transformation of the fundamental spinor representation of $SO(d-1,1)$ can only have non-vanishing matrix elements (3)  for a massless particle with the Lorentz transformation of the fundamental spinor representation of $SO(d-2)$.  The matrix element (5)  (and the free field) necessarily satisfies the zero mass Dirac equation.
  \item A field  with the Lorentz transformation of the direct product of the fundamental spinor representation of $SO(d-1,1)$ and a $d$ vector, satisfying an irreducibility condition, can only have non-vanishing matrix elements (3)  for a massless particle with the Lorentz transformation of the fundamental spinor representation of $SO(d-2)$.  The matrix element (5) (and the free field)  is necessarily the spacetime derivative of a spinor field satisfying the zero mass Dirac equation.
   \end{itemize}
Unfortunately a general statement like that for $d=4$ of the allowed massless particle types for fields that furnish arbitrary representations of the $d$-dimensional Lorentz group is still lacking.  A conjecture will be offered for tensor fields furnishing general irreducible representations of $SO(d-1,1)$, that they can only describe massless particles that transform under representations of $SO(d-2)$ characterized by a Young tableau obtained by decapitating the  Young tableau characterizing the field.

These results appear quite different from those given by  Labastida for several types of tensor and tensor-spinor field [5].
The difference arises from different assumptions.  In the present work, a tensor field is understood to transform strictly as a tensor, while for Labastida it may instead transform as a tensor only up to a gauge transformation.  Labastida therefore sought fields that satisfy gauge-invariant field equations, while here the field equations if any follow directly from the field transformation property and Wigner's requirement that the generators of the invariant Abelian subalgebra of the little group algebra must be represented on physical states of a single massless particle by zero.  

Where gauge transformations do enter here, it is not for the tensor fields, but for their potentials.  For instance, as noted above, an antisymmetric tensor free field of rank $N$ constructed from massless  particle creation and annihilation operators must be the exterior derivative of a potential that is antisymmetric in $N-1$ spacetime indices, but that rank $N-1$ potential does not transform as a tensor, but rather as a tensor up to a gauge transformation that leaves the rank $N$ tensor field invariant.  

This is familiar from the examples of electrodynamics and general relativity in $d=4$ spacetime dimensions.  
From the creation and annihilation operators of a photon of helicity $\pm 1$ we can construct a true antisymmetric tensor field $F^{\mu\nu}$, which satisfies homogeneous Maxwell equations that amount to the statement that its exterior derivative vanishes.  The tensor field $F^{\mu\nu}$
is itself the exterior derivative of a  potential $A^\mu$,  but since this potential is constructed only from the creation and annihilation operators of photons of helicity $\pm 1$, it cannot transform as a four-vector under Lorentz transformations, but rather as a four-vector only up to a gauge transformation, such as the electromagnetic four-vector potential in Coulomb gauge.    
Similarly, from the creation and annihilation operators of a graviton of helicity $\pm 2$ we can construct a linearized Weyl curvature tensor $C_{\mu\nu\rho\sigma}$ that has the Lorentz transformation of a true tensor, which  satisfies linearized Bianchi identities.  The tensor field $C_{\mu\nu\rho\sigma}$ is the second spacetime-derivative of a symmetric  potential $h_{\mu\nu}$,  but since this potential is constructed only from the creation and annihilation operators of gravitons of helicity $\pm 2$, it cannot transform as a true tensor  under Lorentz transformations, but rather as a tensor only up to a gauge transformation, such as the metric perturbation in transverse-traceless gauge.  

From this point of view, gauge invariance is not a fundamental assumption, as it is in Labastida's work, but is rather a consequence of the Lorentz transformation properties of fields and massless particle states.

\begin{center}
{\bf 2. THE LITTLE ALGEBRA}
\end{center}

To derive the results cited above, we will use the fundamental requirement (4) in the limit where 
\begin{equation}
W^\mu{}_\nu\rightarrow\delta^\mu{}_\nu+\Omega^\mu{}_\nu\;,
\end{equation}
with $\Omega^\mu{}_\nu$ infinitesimal and constrained by the conditions that $W^\mu{}_\nu$ should belong to the little group:
\begin{equation}
\Omega_{\mu\nu}=-\Omega_{\nu\mu}\;,~~~~~\Omega^\mu{}_\nu k^\nu=0\;.
\end{equation}
  For such transformations, we write
\begin{equation}
d_{\bar{\sigma},\sigma}(1+\Omega)\rightarrow \delta_{\bar{\sigma},\sigma}+\frac{i}{2}J^{\mu\nu}_{\bar{\sigma},\sigma}\Omega_{\mu\nu}\;,
\end{equation}
with Hermitian matrices $J^{\mu\nu}_{\bar{\sigma},\sigma}=-J^{\nu\mu}_{\bar{\sigma},\sigma}$, satisfying the $SO(d-1,1)$ commutation relations
\begin{equation}
i[J^{\mu\nu},J^{\rho\sigma}]=\eta^{\nu\rho}J^{\mu\sigma}-\eta^{\mu\rho}J^{\nu\sigma}-\eta^{\sigma\mu}J^{\rho\nu}+\eta^{\sigma\nu}J^{\rho\mu}\;.
\end{equation}
The matrices representing the little group generators  are $J^{ij}$, with $i$, $j$, etc. here and below running over the values $1,2,3,\dots d-2$, together with 
\begin{equation}
K^i\equiv J^{i\;d-1}-J^{i0}\;.
\end{equation}
The little group generators have the commutators
\begin{equation}
i[J^{ij},J^{kl}]=\delta_{jk}J^{il}-\delta_{ik}J^{jl}-\delta_{li}J^{kj}+\delta_{lj}J^{ki}\,
\end{equation}
\begin{equation}
i[J^{ij},K^{k}]=\delta_{jk}K^{i}-\delta_{ik}K^{j}\,
\end{equation}
\begin{equation}
[K^ i,K^j]=0\;.
\end{equation}
Because the $K^j$ span an invariant Abelian subalgebra, they have to annihilate $u$:
\begin{equation}
\sum_{\bar{\sigma}} [K^j]_{\sigma,\bar{\sigma}} u_{\bar{\sigma}}^n=0
\end{equation}

For a general tensor field $\psi^{\mu\nu\dots}$, we have
$$[D(1+\Omega)u_\sigma]^{\mu\nu\dots}=(\delta^\mu{}_\kappa+\Omega^\mu{}_\kappa)(\delta^\nu{}_\lambda+\Omega^\nu{}_\lambda)\cdots  u^{\kappa\lambda\dots}_\sigma\;,$$
  It is very convenient  to use light-cone coordinates 
in which for a general vector $V^\mu$,
\begin{equation}
V^\pm\equiv \frac{1}{2}\left[V^0\pm V^{d-1}\right]
\end{equation}
and likewise for the spacetime indices on tensors and spinor-tensors.
By comparing coefficients of $\Omega^\mu{}_\nu$ on both sides of Eq.~(4) we can work out the action on $u$ of the generators $K^j$  in the representation $d_{\bar{\sigma},\sigma}$, so that Eq.~(17) gives:
\begin{eqnarray}
&&0=\sum_{\bar{\sigma}} [K^j]_{\sigma,\bar{\sigma}} u_{\bar{\sigma}}^{+\dots}= iu_{\sigma}^{j\dots}+\dots \;, \\
&& 0= \sum_{\bar{\sigma}} [K^j]_{\sigma,\bar{\sigma}} u_{\bar{\sigma}}^{i\dots}= 2i\delta_{ij}u_{\sigma}^{-\dots}+\dots\;,\\
&&0=\sum_{\bar{\sigma}}[K^j]_{\sigma,\bar{\sigma}} u_{\bar{\sigma}}^{-\dots}=0+\dots\;,
\end{eqnarray}
again with $i$ and $j$ taking the values $1,2,\dots,d-2$.
(The final ``$+\dots$'' in each formula refers to the action of $K^i$ on the undisplayed indices on $u$, indicated in the superscript by $\dots$.)

  For spinors and spinor-tensors we also need to know that for $\chi_n$ in the fundamental spinor representation of $O(d-1,1)$:
\begin{equation}
\sum_s J_{rs}^{\mu\nu}\chi_s=-\frac{i}{4}\sum_s[\Gamma^\mu,\Gamma^\nu]_{rs}\chi_s\;,
\end{equation}
where the $\Gamma^\mu_{rs}$ are $2^{d/2}$-dimensional  matrices forming a Clifford algebra with ( suppressing indices $r,s$)
\begin{equation}
\Gamma^\mu\Gamma^\nu+\Gamma^\nu\Gamma^\mu=2\eta^{\mu\nu}\;,
\end{equation}
and therefore for a general spinor-tensor
\begin{equation}
K^iu^{\dots}=i\Gamma^i\Gamma^-u^{\dots}+\dots\;,
\end{equation}
where the final ``$+\dots$'' denotes the action of $K^i$ on the tensor indices, which are indicated as the superscript ``$\dots$'' on $u$.
From Eqs.~(19)--(21) and (24) we can work out the action of $K^i$ on tensors and spinor-tensors.  

We are now equipped to derive the consequences of the requirement that $K^iu=0$ for fields of various special types.

\begin{center}
{\bf 3.  TENSORS}
\end{center}

We  now derive the results  listed in Section 1 for tensor fields.

\vspace{10pt}

\noindent
{\bf Symmetric traceless tensors}\\
Consider a symmetric traceless tensor field $\psi^{\mu_1\mu_2\dots\mu_N}$  of arbitrary rank $N$.  A general component of the matrix element (3)  with $N_+$ + indices,  $N_-$ $-$ indices, and $M=N-N_+-N_-$ indices in the range of 1 to $d-2$ will be denoted  $u^{i_1 i_2 \dots i_M\;(N_+,N_-)}$ .  By using Eqs.~(19)-(21), we see that the condition $K^ju=0$ applied to such a component gives
\begin{equation}
0=u^{i_1 i_2 \dots i_M j\;(N_+-1,N_-)}+2\sum_{r=1}^M \delta_{ji_r} u^{i_1 i_2 \dots i_{r-1}i_{r+1} \dots i_M\;(N_+,N_-+1)}\;.
\end{equation}
Setting $i_M=j$, summing over $j$, and using the symmetry of $u$  gives
\begin{equation}
0=\sum_{j}u^{i_1 i_2 \dots i_{M-1}j j\;(N_+-1,N_-)}+2M  u^{i_1 i_2 \dots i_{M-1}\;(N_+,N_-+1)}\;.
\end{equation}
In light-cone coordinates the metric tensor has non-vanishing components
\begin{equation}
\eta_{ij}=\delta_{ij}\;,~~~~~~~~~~\eta_{+-}=\eta_{-+}=-2
\end{equation}
so the condition that $u$ is traceless gives
$$\sum_{j}u^{i_1 i_2 \dots i_{M-1}jj\;(N_+-1,N_-)}=4u^{i_1 i_2 \dots i_{M-1}\;(N_+,N_-+1)}$$
and Eq.~(26) therefore reads
$$
 0= (4+2M)  u^{i_1 i_2 \dots i_{M-1}\;(N_+,N_-+1)}\;.
$$
We see that all components $u^{\mu\nu\dots}$ vanish for which any of the indices $\mu$, $\nu$, {\em etc}. is a  $-$.  The second term in Eq.~(25) is therefore absent, and Eq.~(25)  becomes
\begin{equation}
0=u^{i_1 i_2 \dots i_M j\;(N-M-1,0)}\;,
\end{equation}
for any $M\leq N-1$.  The only remaining non-zero component of $u^{\mu\nu\dots}$ is $u^{++\dots}$, with all indices equal to $+$.  

Because $u^{i+++\dots}=0$ this $u$ satisfies the condition $K^iu=0$.  It is also  invariant under the semi-simple part $SO(d-2)$ of the little group, so the only massless particle that can be destroyed by a  symmetric traceless tensor must transform trivially under $SO(d-2)$.  

With $u^{\mu\nu\dots}({\bf k})$  having only  non-zero components with all indices $+$, it is simply proportional to $k^\mu
k^\nu\cdots$, and the  field matrix element then is just the $N$-th partial derivative of a scalar:
\begin{equation}
\langle vac|\psi^{\mu_1\mu_2\dots\mu_N}(x)|{\bf k}\rangle\propto\frac{\partial^N  e^{ik\cdot x}}{\partial x_{\mu_1}\partial x_{\mu_2}\cdots \partial x_{\mu_N}}
\end{equation}
Equivalently, the free field has the form
\begin{equation}
\psi^{\mu_1\mu_2\dots\mu_N}(x)=\frac{\partial^N \varphi(x)}{\partial x_{\mu_1}\partial x_{\mu_2}\cdots \partial x_{\mu_N}}
\end{equation}
where
\begin{equation}
\varphi(x)=\int d^{d-1}p\left[u a({\bf p})e^{ip\cdot x}+
 v b^\dagger({\bf p})e^{-ip\cdot x}\right]\;,
\end{equation}
with $u$ and $v$ some complex numbers.

 \vspace{10pt}

\noindent
{\bf Antisymmetric  tensors}\\
Next consider a completely antisymmetric tensor field $\psi^{\mu_1\mu_2\dots\mu_N}$ of any rank $N\leq d$.  The components of the coefficient function 
$u^{\mu_1\mu_2\dots\mu_N}({\bf k},\sigma)$
 are of four types: those with no $+$ and no $-$ indices and $N$ indices between $1$ and $d-2$;  those with one $+$ and no $-$ indices and $N-1$ indices between $1$ and $d-2$; those with one $-$ and no $+$ indices and $N-1$ indices between $1$ and $d-2$; and those with one $+$ and one $-$ indices and $N-2$ indices between $1$ and $d-2$.  Using Eqs.~(19)--(21), we see that the condition $K^ju=0$  applied to each component type then gives
\begin{equation}
0=K^ju^{i_1 i_2\dots  i_{N}}=2i\sum_{r=1}^N \delta_{ji_r}u^{i_1 i_2\dots i_{r-1}\;-\;i_{r+1}\dots  i_N }\;,
\end{equation} 
\begin{equation}
0=K^ju^{i_1 i_2\dots  i_{N-1}+} =2i\sum_{r=1}^{N-1}\delta_{ji_r}u^{i_1 i_2\dots i_{r-1}\;-\;i_{r+1}\dots i_{N-1}+}  + iu^{i_1\dots i_{N-1}j} \;,
\end{equation}
\begin{equation}
0=K^ju^{i_1\dots i_{N-2}+-}=i u^{i_1\dots i_{N-1}j-}  \;.
\end{equation}
(As a consequence of antisymmetry, $K^ju^{i_1 i_2\dots  i_{N-1}-}$ vanishes automatically.)
Setting  $i_{N-1}=j$ in Eq.~(33)  and summing over $j$  gives
  \begin{equation}
0=u^{i_1 i_2\dots i_{N-2}+-}\;.
\end{equation}
which with Eq.~(34) shows that any component $u^{\mu\nu\dots}$ in which one of the spacetime indices is a $-$ vanishes, whether or not another of the indices is a $+$.  
Hence Eq.~(32) is satisfied, and Eq.~(33)  shows that any component $u^{\mu\nu\dots}$ in which none of the spacetime indices is a $+$ vanishes.  This leaves the only non-vanishing component of the form $u^{i_1 i_2\dots  i_{N-1}+}$.  

Because $u^{i_1 i_2\dots  i_{N}}=0$ and $u^{i_1 i_2\dots  i_{N-2}-+}=0$, this $u$ satisfies the condition $K^iu=0$.  Also, it has the transformation property under $SO(d-2)$ of a completely antisymmetric tensor of rank $N-1$, and this is the transformation of the one-particle states under $SO(d-2)$.  

Since $u^{i_1 i_2\dots  i_{N-1}\mu}$ vanishes except for $\mu=+$, it is proportional to $k^\mu$, and  
$u^{\mu_1\mu_2\dots\mu_N}$ therefore takes the form $u^{[\mu_1\mu_2\dots\mu_{N-1}}k^{\mu_N]}$, with the square brackets indicating antisymmetrization in the indices $\mu_1\mu_2\dots\mu_N$.  The matrix element (5) (and the free field $\psi^{\mu_1\mu_2\dots\mu_N}(x)$) is therefore the exterior derivative of an antisymmetric potential of rank $N-1$.

\vspace{10pt}

\noindent
{\bf Weyl  tensor}\\
We now consider a fourth-rank traceless tensor $\psi^{\mu\nu,\,\rho\sigma}$ with  the symmetry properties of the Weyl tensor:
\begin{equation}
\psi^{\mu\nu,\,\rho\sigma}=-\psi^{\nu\mu,\,\rho\sigma}=-\psi^{\mu\nu,\,\sigma\rho}=+\psi^{\rho\sigma,\,\mu\nu}
\end{equation}
\begin{equation}
\psi^{\mu\nu,\,\rho\sigma}+\psi^{\mu\rho,\,\sigma\nu}+\psi^{\mu\sigma,\,\nu\rho}=0\;.
\end{equation}
(Eq.~(37) has as a  consequence the vanishing of any completely antisymmetric part of $\psi$, which otherwise by itself would satisfy Eq.~(36).)

Our first task will be to show that $u^{\mu\nu,\,\rho\sigma}({\bf k},\sigma)$ vanishes if any  of the indices $\mu,\,\nu,\,\rho,\,\sigma$ is a minus.
First, note that $K^ju^{+-,+-}=2iu^{j-,+-}$ so $u^{j-,+-}=0$.  Also,   $K^ju^{i-,+-}=iu^{i-,j-}$ (because antisymmetry makes $u^{--,+-}$ vanish) so 
$u^{i-,j-}=0$.  Further, 
$$
0=-iK^j u^{+-,+i}=u^{j-,+i}+u^{+-,ji}+2\delta_{ij}u^{+-,+-}\;.
$$
The assumed tracelessness of $u$ gives $\sum_j u^{j-,+j}=2u^{+-,+-}$ and of course $u^{+-,jj}=0$,  so setting $i=j$  and summing over $j$ gives
$$0=2u^{+-,+-}+2(d-2)u^{+-,+-}\;,$$
and so $u^{+-,+-}=0$.  We have thus shown that $u^{\mu\nu,\,\rho\sigma}({\bf k},\sigma)$ vanishes if any {\em two} of the indices $\mu,\,\nu,\,\rho,\,\sigma$ are minus.  

To consider the case where just one of the indices $\mu,\,\nu,\,\rho,\,\sigma$ is a minus, first note that because $u^{\mu\nu,\,\rho\sigma}({\bf k},\sigma)$ vanishes if any two of its indices is a minus, we have $K^iu^{jk,+-}=iu^{jk,i-}$ so $$u^{jk,i-}=0\;.$$  On the other hand,
\begin{equation}
0=-iK^iu^{j+,+-}=u^{ji,+-}+u^{j+,i-}\;.
\end{equation}
We can also show that a different linear combination of the two terms in Eq.~(38) vanishes.  For this purpose, we use
$$ 0=-iK^i u^{jk,l+}=2\delta_{ij}u^{-k,l+}+2\delta_{ik}u^{j-,l+}+2\delta_{il}u^{jk,-+}+u^{jk,li}\;.$$
Setting $i=j$ and summing over $j$ gives
$$0=2(d-2)u^{-k,l+}+2u^{k-,l+}+2u^{lk,-+}+2 u^{+k,l-}+2u^{-k,l+}$$
If we add to this the same equation with $l$ and $k$ interchanged we find
$$u^{-k,l+}+u^{-l,k+}=0$$
so 
$$ 2(d-3)u^{-k,l+}+2u^{lk,-+}=0\;.$$
Comparing this with Eq.~(38) (with $i$ and $j$ replaced with $k$ and $l$) we see that for $d\neq 2$ both terms vanish:
$$u^{-k,l+}=0\;,~~~~u^{lk,-+}=0\;.$$
Further, we use
$$0=-iK^iu^{j+,k+}=2\delta_{ij}u^{-+,k+}+u^{ji,k+}+2\delta_{ik}u^{j+,-+}+u^{j+,ki}\;.$$
Setting $i=j$ and summing over $j$ this gives
$$
0=2(d-2)u^{-+,k+}+2u^{k+,-+}+2u^{-+,k+}=2du^{-+,k+}\;,$$
so also $u^{-+,k+}=0$.
Thus as  promised, we have shown that $u^{\mu\nu,\,\rho\sigma}({\bf k},\sigma)$ vanishes if any  of the indices $\mu,\,\nu,\,\rho,\,\sigma$ is a minus.

It follows that $K^iu^{jk,l+}=iu^{jk,li}$ so $u^{jk,li}=0$.  This leaves $u^{ij,k+}$ and $u^{i+,j+}$ as the only independent components of $u$ that have not been shown to vanish.  Furthermore, with all other components vanishing we have $K^iu^{jk,l+}=0$ and  
$$0=-iK^i u^{j+,k+}=u^{ji,k+}+u^{j+,ki}=u^{ji,k+}+u^{ki,j+}$$
so $u^{ji,k+}$ is antisymmetric in $i$, $j$, and $k$, and therefore vanishes as a consequence of Eq.~(37). The only remaining non-zero component $u^{i+,j+}$ transforms under $SO(d-2)$ as a symmetric traceless second-rank tensor, as was to be shown.

Taking account of the symmetry assumption (36),  this   takes the form
$$u^{\mu\nu,\rho\sigma}=k^\mu k^\rho a^{\nu\sigma}-k^\nu k^\rho a^{\mu\sigma}-k^\mu k^\sigma a^{\nu\rho}+k^\nu k^\sigma a^{\mu\rho}\;,$$
where $a^{\mu\nu}$ is symmetric  and, because all components with a minus index vanish, also traceless:
$$ k_\mu a^{\mu\nu}=0\;,\;.$$  The free  field thus takes the form
\begin{equation}
\psi^{\mu\nu,\rho\sigma}(x)=\partial^\mu \partial^\rho h^{\nu\sigma}(x)-\partial^\nu \partial^\rho h^{\mu\sigma}(x)-\partial^\mu \partial^\sigma h^{\nu\rho}(x)+\partial^\nu \partial^\sigma h^{\mu\rho}(x)
\end{equation}
where $h^{\mu\nu}$ is a symmetric potential.

\begin{center}
{\bf 4.  SPINORS AND SPINOR-VECTORS}
\end{center}

We  now derive the results  listed in Section 1 for spinor and spinor-vector fields.

\vspace{10pt}

\noindent
{\bf Spinors}\\
We first consider a field $\psi_n$ that transforms according to the fundamental spinor representation of $SO(d-1,1)$.  From Eq.~(24) we see that the 
requirement that $K^iu=0$ gives $\Gamma^i\Gamma^-u=0$, and since $(\Gamma^i)^2=1$, this requires that $\Gamma^-u=0$, or in other words
\begin{equation}
\Gamma^\mu k_\mu u=0\;.
\end{equation}
  So, since the generators $-i[\Gamma_i,\Gamma_j]$ of the $SO(d-2)$ subgroup of the little group commute with $\Gamma^\mu k_\mu$, and there are no other conditions on $u$, the massless particle  states described by this field transform according to the fundamental spinor representation of $SO(d-2)$.  Also, it follows from Eq.~(40) that the free field must satisfy the massless Dirac equation
\begin{equation}
\Gamma^\mu\frac{\partial}{\partial x^\mu}\psi(x)=0\;.
\end{equation}

\vspace{10pt}

\noindent
{\bf Spinor-Vectors}\\
We next consider a field $\psi^\mu$ that transforms according to the direct product of the vector and fundamental spinor representations of $SO(d-1,1)$,
and is irreducible, in the sense that
\begin{equation}
\eta_{\mu\nu}\Gamma^\mu\psi^\nu=0
\end{equation}
so that it is not possible to separate out a part of $\psi^\mu$ that transforms simply according to the fundamental spinor representation of $SO(d-1,1)$.
  By combining  Eqs.~(19)--(21) and (24) we see that the coefficient functions  $u^\mu({\bf k},\sigma)$ must satisfy
\begin{equation}
0=K^ju^i=2i\delta_{ij}u^--i\Gamma^j\Gamma^-u^i\;,
\end{equation}
\begin{equation}
0=K^ju^+=iu^j -i\Gamma^j\Gamma^-u^+\;,
\end{equation}
\begin{equation}
0=K^ju^-=-i\Gamma^j\Gamma^-u^-\;.
\end{equation}
From Eqs.~(44) and (45) it follows that
$$
2\delta_{ij} u^-=\Gamma^j\Gamma^-\Gamma^i\Gamma^-u^+=-\Gamma^j\Gamma^i(\Gamma^-)^2u^+\;.
$$
This vanishes because $(\Gamma^-)^2=0$, so 
\begin{equation}
u^-=0\;,
\end{equation}
and Eq.~(46) is then automatically satisfied.
Since $(\Gamma_i)^2=1$,  Eq.~(44) then gives
\begin{equation}
\Gamma^- u^i=0\;.
\end{equation}
We next invoke the irreducibility condition (43).  Using Eqs.~(27) and (47), this gives
$$ 0=\sum_i \Gamma^i u^i-2\Gamma^- u^+\;.$$
Multiplying Eq.~(45) with $\Gamma^j$ and summing over $j$ we then have
$$(d-2)\Gamma^-u^+=\sum_j\Gamma^ju^j=2\Gamma^-u^+$$
so, for any $d\neq 4$,
$$
\Gamma^-u^+=0\;,
$$
and Eq.~(45) then gives
\begin{equation}
u^j=0\;.
\end{equation}
Since the only non-vanishing component of $u^\mu$ is $u^+$, the particle states transform according to the fundamental spinor representation of $SO(d-2)$.  Further,  because $u^\mu$ is proportional to $k^\mu$, the vector-spinor free field $\psi^\mu$ is the spacetime derivative $\partial \psi/\partial x_\mu$ of a spinor field $\psi$.

\begin{center}
{\bf 5.  DECAPITATION CONJECTURE}
\end{center}

General tensor fields are characterized by their symmetry properties, symbolized by Young tableaux, along with conditions of tracelessness and in some cases prescriptions on contraction with the $d$-dimensional Levi-Civita symbol.  Briefly, a tensor whose symmetry properties are described by a given Young tableau will be antisymmetric in the spacetime indices associated with the boxes in any column, symmetric under the interchange of any two columns of the same height, and will vanish when antisymmetrized in all the indices in any pair of columns.  In all the cases of tensor fields we have considered, it turned out that the only non-vanishing components of $u_\sigma^{\mu\nu\dots}$ have no $-$ indices.  Further, it turned out that the non-vanishing components had just one $+$ index in each column of the Young tableau that characterizes the symmetries among these indices.  Because a tensor is antisymmetric in the indices in each column, this $+$ index can always be taken to be at the top of the column.  The non-vanishing components of $u_\sigma^{\mu\nu\dots}$ thus have a top row with all $+$ indices, and all other indices below the top row running over the values $1,2,\dots, d-2$.  Since $SO(d-2)$ rotations only act on these lower indices, we are led to the conjecture that if a field furnishes an irreducible tensor  representation of $SO(d-1,1)$ with symmetry dictated by some Young tableau, then the massless particle states form a tensor representation of the semi-simple part $SO(d-2)$ of the little group with symmetry dictated by the same Young tableau, but decapitated by deleting the top row.  

For instance, in the case of a symmetric traceless tensor of rank $N$ the Young tableau has just a single row of width $N$, which when deleted according to this conjecture leaves massless particle states invariant under $SO(d-2)$, as we found here.
In the case of an antisymmetric tensor of rank $N$ the Young tableau has just a single column of height $N$, which when the top box is deleted according to this conjecture leaves massless particle states that transform under $SO(d-2)$ as an antisymmetric tensor of rank $N-1$, also as we found here.
In the case of a Weyl tensor, a traceless tensor of rank 4 with a $2\times 2$ Young tableau, when the top row is deleted we are left with a single row of width 2, which according to this conjecture indicates that the field describes massless particles that transform as a traceless symmetric tensor of rank 2, again as we found here.  

Furthermore,  it is easy to see that this conjecture works in the case $d=4$ for traceless tensor fields with symmetry described by arbitrary Young tableaux.  With $d=4$ it is not possible to have  Young tableaux with columns of height greater than 4, while by using the Levi-Civita symbol columns of height 3 or 4 can be replaced with columns of height 1 or 0, so the general  Young tableaux have columns only of height 0, 1 and 2.  A traceless tensor with a Young tableau having $n_1$ columns of height 1 and $n_2$ columns of height 2 has Lorentz transformation type
$$ (n_2+n_1/2, n_1/2)\oplus (n_1/2, n_2+n_1/2)\;,$$ so it can only describe massless particles of helicity (8) equal to $\pm n_2$.  On the other hand, removing the top row of a Young tableau having $n_1$ columns of height 1 and $n_2$ columns of height 2 yields a Young tableau with just one row of width $n_2$, so the massless particles described by such a field transform under $SO(2)$ as a symmetric tensor of rank $n_2$, and therefore indeed have helicity $\pm n_2$.  

Unfortunately the methods used in this paper are too inefficient to lead easily to a general proof of this conjecture.

\vspace{10pt}

\vspace{10pt}

\begin{center}
{\bf 	ACKNOWLEDGMENTS}
\end{center}

I am grateful to Jacques Distler for information about Young tableaux.  Thanks also to Yale Fan for useful comments.  This work  is supported by the National Science Foundation under grant number
Phy-1914679 and also with support from the Robert A. Welch Foundation, Grant No. F-0014. 

\begin{center}
-------------------------------
\end{center}

\begin{description} 
\item[1] T, Kaluza, Sitz. Preuss. Akad.  Wiss. {\bf K1}, 966 (1921); O. Klein, Z. Phys. {\bf 37}, 895 (1926); Nature {\bf 118}, 515 (1926).
\item[2]  E. P. Wigner, Ann. Math. {\bf 40}, 149 (1939).
\item[3] N. Ohta, Phys. Rev. D {\bf 31}, 442 (1985), has shown that with a suitable normalization of $u$ and $v$ the fields constructed in this way commute with each other outside the light cone if the particles they describe are bosons for tensor fields and fermions for spinor-tensor fields.
\item[4] S. Weinberg, Phys. Rev. {\bf 134}, B882 (1964).  For a textbook account see S. Weinberg, {\em The Quantum Theory of Fields}, Vol I: Foundations (Cambridge University  Press, Cambridge, UK, 1995), Section 5.9.
\item[5] J. M. F. Labastida, Nucl. Phys. B {\bf 322}, 185 (1989). 
\end{description}

\end{document}